\documentclass[  reprint,  showpacs,  showthanks, onecolumn, secnumarabic,
amsmath, amssymb,  aps, pra, nobibnotes, floats, superscriptaddress]{revtex4}%
\usepackage{amsmath}
\usepackage{graphicx}
\usepackage{amsfonts}
\usepackage{amssymb}%
\providecommand{\U}[1]{\protect\rule{.1in}{.1in}}
\providecommand{\U}[1]{\protect\rule{.1in}{.1in}}
\newtheorem{theorem}{Theorem}
\newtheorem{acknowledgement}[theorem]{Acknowledgement}

\begin{document}
\title{{\Large Peculiarities of electron energy spectrum in Coulomb field of super
heavy nucleus}}
\author{D.M. Gitman}
\affiliation{Division of Theoretical Physics, P.N. Lebedev Physical Institute, Russia;
Tomsk State, University, Russia; Institute of Physics, University of S\~{a}o
Paulo, Brazil}
\email{gitman@if.usp.br}
\author{B.L. Voronov}
\affiliation{Division of Theoretical Physics, P.N. Lebedev Physical Institute, Russia}
\email{voronov@lpi.ru}
\author{R. Ferreira}
\affiliation{Institute of Physics, University of S\~ao Paulo, Brazil}
\email{rafaelufpi@gmail.com}
\author{A.D. Levin}
\affiliation{Institute of Physics, University of S\~ao Paulo, Brazil}
\email{Alexander.D.Levin@gmail.com}

\begin{abstract}
Just after the Dirac equation was established, a number of physicists tried to
comment on and solve the spectral problem for the Dirac Hamiltonian with the
Coulomb field of arbitrarily large charge $Z$, especially with $Z$ that is
more than the critical value $Z_{\mathrm{c}}=\alpha^{-1}\simeq137,04$, making
sometimes contradictory conclusions and presenting doubtful solutions. It
seems that there is no consesus on this problem up until now and especially on
the way of using corresponding solutions of the Dirac equation in calculating
physical processes. That is why in the present article, we turn once again to
discussing peculiarities of electron energy spectrum in the Coulomb field of
superheavy nucleus. In the beginning, we remind the reader of a long story
with a wrong interpretation of the problem in the case of a point nucleus and
its present correct solution. We then turn to the spectral problem in the case
of a regularized Coulomb field. Under a specific regularization, we derive an
exact spectrum equation determining the point spectrum in the energy interval
$(-m,m)$ and present some of its numerical solutions. We also derive an exact
equation for charges $Z$ providing bound states with energy $E=-m$. Its
analytical and numerical analysis shows that there exists an infinite number
of such charges; in this connection , we discuss the notion of supercritical charge.

To our mind, their existence does not mean that the one-particle relativistic
quantum mechanics based on the Dirac Hamiltonian with the Coulomb field of
such charges is mathematically inconsistent. In any case, it is physically
unacceptable because the spectrum of the Hamiltonian is unbounded from below,
which requires the secondary Fermi--Dirac quantization and transition to
many-particle quantum field theory. The consequences of the existence of such
charges for quantum electrodynamics with the corresponding Coulomb field
remain to be established in the process of constructing such a theory.

\end{abstract}

\maketitle

\begin{center}
\textbf{DG and BV devote this paper to their friend and permanent coauthor
Igor Tyutin}
\end{center}

\section{Introduction\label{S1}}

Relativistic quantum effects, in particular, electron-positron pair creation,
in superstrong Coulomb field attract attention of physicists already for a
long time. However, their qualitative and especially quantitative description
is lacking up until now. We believe that such a description \ can be done only
in the framework of a nonperturbative quantum electrodynamics (QED) with
superstrong Coulomb field as an external background. Unfortunately, such a
version of QED does not exist at present. Our experience in quantum field
theory (QFT) with different backgrounds, see Refs. \cite{Gitman,277}, allows
us to expect that constructing this version of QED needs at least a complete
and clear mathematical solution of the spectral problem for the Dirac
Hamiltonian with the Coulomb field of arbitrarily large charge $Ze\ $($e>0$ is
the absolute value of the electron charge)\ of a nucleus\textrm{ }(in what
follows, we call $Z$ simply the charge of the nucleus). We also realize that a
solution of the latter problem marks only the beginning of constructing QED
with superstrong Coulomb field. It should be noted that just after the Dirac
equation was established, a number of physicists tried to comment on and solve
this spectral problem making sometimes contradictory conclusions. It seems
that there is no consesus on this problem even at present. That is why we turn
once again to a discussion of peculiarities of the energy spectrum of an
electron in the Coulomb field of a superheavy nucleus. The paper is organized
as follows. In Sec.\ref{S2}, we recall a long story with controversial
interpretations of the problem in the case of a point nucleus and its present
correct solution. Then in Sec. \ref{S3}, we turn to the spectral problem in
the case of a regularized Coulomb field with a specific cutoff, which allows
an exact solution. We analyze the problem in the part concerning the point
(discrete) spectrum located in the segment $[-m,m]$ and the corresponding
bound states. In contrast to the earlier works, we do not use the
approximation of small cutoff radius. In Sec. \ref{S4}, we derive an exact
spectrum equation determining the point spectrum in the energy interval
$(-m,m)$ and present some of its numerical solutions related to different $Z$.
In Sec. \ref{S5}, we derive exact equations for the charges providing the
bound states with energies $E=-m$ and show that there exists an infinite
number of such charges, generally not integer-valued; the first of these
charges are calculated numerically. In this connection, we discuss a
controversial notion of supercritical charge.

\section{Spectral problem with Coulomb field of point nucleus\label{S2}}

The spectral problem for the Dirac Hamiltonian with the Coulomb field of a
point nucleus has a long story. The electronic structure of an atom with
$Z\leq Z_{\mathrm{c}}=\alpha^{-1}\simeq137,04,$ where $\alpha$ is the fine
structure constant, and $Z_{\mathrm{c}}$ is the critical charge, was described
by the Dirac equation, which gives relativistic electron energy spectrum (the
Sommerfeld spectrum) in agreement with experiment~\cite{BetSa57}. It was
commonly believed that the Dirac equation with nucleus charges
$Z>Z_{\mathrm{c}}$ meets insuperable difficulties
\cite{Dirac28,Rose61,AkhBe69,Grein85}. However a short time ago, it was
demonstrated that the common belief that the Dirac Hamiltonian with the
Coulomb field of a point nucleus is consistent only at $Z<Z_{\mathrm{c}}$ is
erroneous{\large ,} see \cite{VorGiT07,book,GitLeTV13}. The known difficulties
with its spectrum for $Z>Z_{\mathrm{c}}$ do not arise if the Dirac Hamiltonian
$\hat{H}(Z)$ is correctly defined as a self-adjoint (s.a.) operator (A first
heuristic attempt in this direction is due to \cite{Case50}). An important
remark concerning admissible values of charge $Z$ is appropriate here. Only
integer-valued $Z$, $Z\in\mathbb{N}$, have a physical meaning, but from the
standpoint of the spectral analysis of the Dirac Hamiltonian, it is useful,
and is commonly adopted, to consider $Z$ as a parameter taking arbitrary
nonnegative values, $Z\in\mathbb{R}_{+}$.

It was demonstrated that from a mathematical standpoint, a definition of the
Dirac Hamiltonian as a s.a. operator presents no problem for arbitrary $Z$.
The Dirac Hamiltonian $\hat{H}\left(  Z\right)  $ with any $Z$ can be
correctly defined as a s.a. operator in the Hilbert space of bispinors.

For $Z<Z_{\mathrm{s}}=\left(  \sqrt{3}/2\right)  \alpha^{-1}\simeq118,7$,
where $Z_{\mathrm{s}}$ is the lower critical charge, the\ Dirac Hamiltonian
$\hat{H}\left(  Z\right)  $ is defined uniquely. For $Z\geq Z_{\mathrm{s}},$
there exists a family $\{\hat{H}^{\left(  \nu\right)  }\left(  Z\right)  \}$
of possible s.a. Dirac Hamiltonians $\hat{H}^{\left(  \nu\right)  }\left(
Z\right)  $ specified by additional boundary conditions at the origin, $\nu$
is generally a certain $Z$ dependent set of parameters. The spectrum and
inversion formulas were found for any $\hat{H}^{\left(  \nu\right)  }\left(
Z\right)  $. The eigenfunctions of the discrete spectrum and generalized
eigenfunctions of the continuous spectrum form a complete orthonormalized
system in the Hilbert space of bispinors. The continuous spectrum is the union
of the two semiaxis $E\leq-m$ and $E\geq m$, while the discrete spectrum
$\{E_{n}^{\left(  \nu\right)  }\left(  Z\right)  \}$ is located in the
interval $|E|\leq m$. The position of discrete energy levels $E_{n}^{\left(
\nu\right)  }\left(  Z\right)  $ essentially depends on $\nu$, in particular,
for any $Z\geq Z_{\mathrm{s}}$, there exist parameters $\nu=\nu_{-m}$, for
which the lowest energy level coincides with the upper boundary $-m$ of the
negative branch $(-\infty,-m]$ of the continuous spectrum,$\ E_{0}^{\nu_{-m}%
}\left(  Z\right)  =-m.$ For $Z<Z_{\mathrm{s}}$, the Sommerfeld spectrum is
generated by the Dirac Hamiltonian $\hat{H}\left(  Z\right)  ,$ while for
$Z_{\mathrm{c}}>Z\geq Z_{\mathrm{s}},$ it is generated by the Dirac
Hamiltonian $\hat{H}^{\left(  0\right)  }\left(  Z\right)  ,$ see \cite{book}.

There is a good reason to believe that these s.a. Dirac Hamiltonians provide
an initial mathematical tool for constructing QED with external strong Coulomb
field of a point charge. The question is how to use this tool and does such
QED exist in principle.

Usually when constructing QFT with an external background, we decompose the
Heisenberg operator of the Dirac field into an adequately chosen complete set
of solutions of the Dirac equation. Our previous experience tells us that to
have a secondly quantized formulation in terms of relatively stable
quasiparticles, the gap between the lowest discrete energy level and the upper
boundary $-m$ of the negative branch of continuous spectrum has to be big
enough. In other words, the discrete energy spectrum has to be isolated enough
from the negative branch $(-\infty,-m]$ of the continuous spectrum. Otherwise,
a desirable secondly quantized theory cannot be constructed in full analogy
with already known cases \cite{Gitman,277}. At least, it is very likely that
such a construction is impossible for s.a. Dirac Hamiltonians $\hat
{H}^{\left(  \nu\right)  }\left(  Z\right)  $ with parameters $\nu=\nu_{-m}.$

\section{Spectral problem with regularized Coulomb field\label{S3}}

\subsection{General}

Before the works \cite{VorGiT07,book,GitLeTV13}, the difficulties with the
energy spectrum of an electron in the Coulomb field of a point nucleus, and
with the spectral problem in general, were explained by a strong singularity
at the origin of the Coulomb field of a nucleus with $Z>Z_{\mathrm{c}}$, see
\cite{Dirac28,Rose61,AkhBe69,Grein85} and many other articles and books. It
was believed that these difficulties can be eliminated if a nucleus of some
finite radius $r_{0}$ is considered. Some calculations were done in support of
the conjecture that with cutting off the Coulomb potential at a finite small
radius $r_{0}$, the Dirac Hamiltonian has a physically meaningful spectrum for
charges $Z$ not exceeding the so-called supercritical charge $Z_{\mathrm{scr}%
}$. Its value depends on the cutoff model and approximations made for its
evaluation. Mention can be made of the following values of the supercritical
charges: $Z_{\mathrm{scr}}=200$ (\cite{PomSm45}), $Z_{\mathrm{scr}}=170$
(\cite{ZelPo72,Popov70}), $Z_{\mathrm{scr}}=172$ (\cite{Grein85}), and some
other values from the interval ($170-177$).

According to the above-listed authors even in the presence of a cutoff, the
lowest discrete level passes into the lower continuum for $Z\geq
Z_{\mathrm{scr}}$. And again the applicability of the Dirac equation now for
nonpoint nuclei with charges $Z\geq Z_{\mathrm{scr}}$ was called into
question. It was supposed that the new difficulties are connected with a
many-particle character of the problem under consideration, in particular,
with a possible $e^{+}e^{-}$ pair creation by a nucleus with the charge $Z\geq
Z_{\mathrm{scr}}$. Since that time almost all researchers in this area
repeated this point of view in their publications. However, recently, there
appeared a publication \cite{UPN} where this conclusion was recognized to be
wrong. For us, after a rehabilitation of the electron spectrum in the Coulomb
field of a point nucleus, it would be very strange to accept the fact that a
removal of the singularity of the Coulomb potential at the origin (after a
cutoff) makes a situation with the spectrum not better, but worse. In view of
a great importance of all the details of the spectral problem for constructing
a corresponding secondly quantized theory, we turn to the spectral problem
with a cutoff once again. We consider this paper as a natural continuation of
our previous works \cite{VorGiT07,book,GitLeTV13}. In this paper, we present
an analytical and numerical study of the spectral problem with a specific
cutoff, under which the problem allows an exact analytical solution. The
presentation is organized as follows. We first briefly recall a reduction of
the general spectral problem to the corresponding radial spectral problem. We
then solve the latter problem partly, namely, in the part concerning the point
spectrum located in the interval $[-m,m]$ and the corresponding bound states,
with minimum references to the previous works on the subject. We plan to
present a detailed comparison of our approach and results with those of
numerous previous papers in a subsequent publication.

\subsection{Radial equations}

Recall that a behavior of an electron in a regularized Coulomb field of charge
$Z$ is governed by the Dirac Hamiltonian $\hat{H}\left(  Z\right)  $ acting in
the Hilbert space $\mathfrak{H=}%
%TCIMACRO{\dsum \limits_{\alpha=1}^{4}}%
%BeginExpansion
{\displaystyle\sum\limits_{\alpha=1}^{4}}
%EndExpansion
\oplus\mathfrak{H}_{\alpha}$, $\mathfrak{H}_{\alpha}=L^{2}(\mathbb{R}^{3})$,
of square-integrable bispinors $\Psi\left(  \mathbf{r}\right)  =\{\psi
_{\alpha}(\mathbf{r}),\alpha=1,2,3,4\}$ by the Dirac differential operation%
\begin{equation}
\check{H}\left(  Z\right)  =\gamma^{0}\left(  \boldsymbol{\gamma}%
\mathbf{\hat{p}}+m\right)  +V(r),\ \mathbf{\hat{p}}=-i\mathbf{\nabla
,\ }r=\left\vert \mathbf{r}\right\vert , \label{s1}%
\end{equation}
where $V(r)$ is the potential energy of the electron in the regularized
Coulomb field (it is supposed to be spherically symmetric, bounded, and real
valued) and $\gamma^{\mu}$ are the Dirac gamma matrices.

The Hamiltonian $\hat{H}(Z)$ with any $Z$ is a uniquely defined s.a. operator
in $\mathfrak{H}$ because it is a sum of the uniquely defined s.a. free Dirac
Hamiltonian and the bounded s.a. operator of multiplication by the bounded
real-valued function $V(r)$: an addition of a bounded s.a. operator to any
s.a. operator yields a new s.a. operator with the same domain. In contrast to
this, in the case of the nonregularized Coulomb field of a point nucleus,
where a potential is an unbounded operator, a s.a. Dirac Hamiltonian is
defined nonuniquely for $Z>Z_{\mathrm{s}}=(\sqrt{3}/2)\alpha^{-1}$
$\simeq118,7$, and the nonuniqueness is growing with increasing $Z$, see
\cite{VorGiT07,book,GitLeTV13}.

The stationary Schr\"{o}dinger equation $\hat{H}\left(  Z\right)  \Psi\left(
\mathbf{r}\right)  =E\Psi\left(  \mathbf{r}\right)  $ defines the point energy
spectrum of the electron, which is the main subject of our interest.

Choosing solutions of the stationary Schr\"{o}dinger equation in the
well-known form%
\begin{equation}
\Psi_{j,M,\zeta}\left(  \mathbf{r}\right)  =\frac{1}{r}\left(
\begin{array}
[c]{l}%
\Omega_{j,M,\zeta}(\theta,\varphi)f\left(  r\right)  \\
i\Omega_{j,M,-\zeta}(\theta,\varphi)g\left(  r\right)
\end{array}
\right)  ,\label{s.3}%
\end{equation}
where $\Omega_{j,M,\zeta}$ are the normalized spherical spinors, so that
bispinors $\Psi_{j,M,\zeta}$ are common eigenvectors of three commuting s.a.
operators $\boldsymbol{\hat{J}}^{2}$, $\hat{J}_{z}$, and $\hat{K}$, where
$\boldsymbol{\hat{J}}$ is the total\ angular\ momentum and $\hat{K}$ is the
so-called spin operator,%
\begin{align}
&  \boldsymbol{\hat{J}}^{2}\Psi=j(j+1)\Psi,\;\hat{J}_{z}\Psi=M\Psi,\ \hat
{K}\Psi=-\varkappa\Psi\ ,\ \varkappa=\zeta(j+1/2),\nonumber\\
&  \boldsymbol{\hat{J}}=\boldsymbol{\hat{L}}+\mathbf{\Sigma/}%
2,\ \boldsymbol{\hat{L}}=\left[  \mathbf{r}\times\mathbf{\hat{p}}\right]
,\ \hat{K}=\gamma^{0}\left[  1+\left(  \boldsymbol{\Sigma\hat{L}}\right)
\right]  ,\ \label{s.4}%
\end{align}
and $j=1/2,3/2,...$,\ $M=-j,-j+1,...,j$,\ $\zeta=\pm1$, we reduce the above
equation to the radial Schr\"{o}dinger equations%
\begin{equation}
\hat{h}\left(  Z,j,\zeta\right)  F\left(  r\right)  =E(Z,j,\zeta)F\left(
r\right)  ,\ F\in\mathbb{L}^{2}(\mathbb{R}_{+})=L^{2}(\mathbb{R}_{+})\oplus
L^{2}(\mathbb{R}_{+}),\label{s.5}%
\end{equation}
where $\hat{h}\left(  Z,j,\zeta\right)  $ are s.a. partial\textrm{ }radial
Hamiltonians acting in the Hilbert space $\mathbb{L}^{2}(\mathbb{R}_{+})$ of
doublets $F\left(  r\right)  ,$%
\begin{equation}
F(r)=\left(
\begin{array}
[c]{c}%
f(r)\\
g(r)
\end{array}
\right)  \label{s.6}%
\end{equation}
by the radial differential operations%
\begin{equation}
\check{h}(Z,j,\zeta)=-i\sigma_{2}\frac{d}{dr}+\varkappa r^{-1}\sigma
_{1}+V(r)+m\sigma_{3},\label{s.7}%
\end{equation}
where $\sigma_{k},\ k=1,2,3$, are the Pauli matrices,\ see
\cite{Rose61,AkhBe69} and \cite{VorGiT07,book}. The domain $D_{\text{ }}$of
each of the operators $\hat{h}$ consists of doublets $F(r)$ that are
absolutely continuous on $(0,\infty)$, \ are vanishing at zero, $f(0)=g(0)=0$,
and are square integrable together\ with $\check{h}F(r)$ on $(0,\infty)$
(actually, at infinity). This is the so-called natural domain for $\check{h}$,
see \cite{book}.

Because the potential $V(r)$ vanishes at infinity, the spectrum of each of
$\hat{h}$ consists of a continuous part that is the union $(-\infty
,-m]\cup\lbrack m,\infty)$\ of two semiaxis, negative and positive, and a
point spectrum $\{E_{n}(Z,j,\zeta),\ n\in\mathbb{Z}_{+}\}$ located in the
segment $[-m,m]$. The total point spectrum of the Dirac Hamiltonian $\hat
{H}(Z)$ is the union of partial point \ spectra of the radial Hamiltonians
$\hat{h}(Z,j,\zeta).$

The radial Schr\"{o}dinger equation (\ref{s.5}) with fixed $Z,j,\zeta$ implies
the system of equations for the radial functions $f$ and $g$:%
\begin{align}
&  \frac{df(r)}{dr}+\frac{\varkappa}{r}f(r)-k_{+}\left(  r\right)
g(r)=0\Rightarrow g(r)=\frac{1}{k_{+}\left(  r\right)  }\left[  \frac
{df(r)}{dr}+\frac{\varkappa}{r}~f(r)\right]  ,\nonumber\\
&  \frac{dg(r)}{dr}-\frac{\varkappa}{r}g(r)+k_{-}\left(  r\right)
f(r)=0,\ k_{\pm}\left(  r\right)  =E-V\left(  r\right)  \pm m. \label{s.9}%
\end{align}

In what follows, we consider the regularized Coulomb potential of the form%
\begin{equation}
V\left(  r\right)  =-q\left\{
\begin{array}
[c]{c}%
r_{0}^{-1},\ r\leq r_{0}\\
r^{-1},\ r\geq r_{0}%
\end{array}
\right.  ,\ \ q=Z\alpha\ . \label{s.2}%
\end{equation}
It corresponds to the field of the positive charge $Ze$ distributed uniformly
on a nucleus spherical surface of radius $r_{0}$, see FIG.1.%
%TCIMACRO{\FRAME{ftbpFU}{2.3393in}{1.4512in}{0pt}{\Qcb{Regularized Coulomb
%potential.}}{}{fig1.eps}{\special{ language "Scientific Word";
%type "GRAPHIC";  maintain-aspect-ratio TRUE;  display "USEDEF";
%valid_file "F";  width 2.3393in;  height 1.4512in;  depth 0pt;
%original-width 3.6115in;  original-height 2.2312in;  cropleft "0";
%filename 'fig1.eps';file-properties "NPEU";}} }%
%BeginExpansion
\begin{figure}[ptb]%
\centering
\includegraphics[
height=1.4512in,
width=2.3393in
]%
{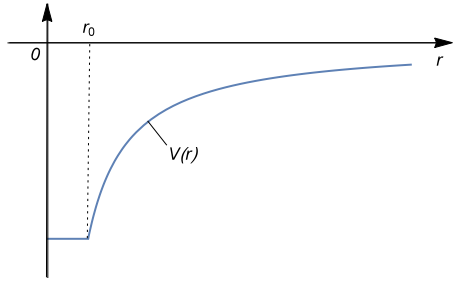}%
\caption{Regularized Coulomb potential.}%
\end{figure}
%EndExpansion

The cutoff radius $r_{0}$ is usually considered a universal $Z$ independent
constant which defines a model. But in accordance with real nuclear physics,
it is natural to consider $r_{0\text{ }}$as a $Z$ dependent parameter,
$r_{0}=r_{0}(Z)$. Under the approximation that the number of protons and
neutrons in a nucleus are equal, this $Z$ dependence is given by
\begin{equation}
r_{0}=r_{0}(Z)=R_{0}(2.5Z)^{1/3},\ \ R_{0}=1.25\times10^{-15}\mathrm{m}%
=0.635\times10^{-8}eV^{-1}, \label{s.18}%
\end{equation}
see \cite{Krane87}. It should be noted that this approximation becomes more
and more rough with increazing $Z$.

In formulas to follow, we write simply $r_{0}\ $for breavity, which allows
applying the formulas to any $r_{0}$, but in numerical calculations, we use
(\ref{s.18}).

In finding point spectra $\{E_{n}(Z,j,\zeta)\}$, we have to consider the open
energy interval $-m<E<m$ and its end points $E=m$ and $E=-m$ separately by
technical reasons explained below in the beginning of Sec. \ref{S5}.

\section{Open energy interval $(-m,m)$, spectrum equation\label{S4}}

\subsection{Solving radial equations in region $0\leq r\leq r_{0}$}

In the internal region $0\leq r\leq r_{0}$, where we set $f(r)=f_{\mathrm{in}%
}(r)$ and$~g(r)=g_{\mathrm{in}}(r)$, the functions $k_{\pm}(r)$ in (\ref{s.9})
become constants. The substitution of the representation for the function
$g_{in}(r)$ from the first row in (\ref{s.9}) into the second equation in
(\ref{s.9}) then results in the following second-order differential equation
for the function $~f_{in}(r)$:%
\begin{align}
&  \frac{d^{2}f_{\mathrm{in}}(r)}{d^{2}r}+\left(  \eta^{2}-\frac{\nu^{2}%
-1/4}{r^{2}}\right)  f_{\mathrm{in}}(r)=0,\ \nonumber\\
&  \eta=\sqrt{k_{+}k_{-}},\ k_{\pm}=E\pm m+\frac{q}{r_{0}},~\nu=j+\frac
{\zeta+1}{2}=\left\{
\begin{array}
[c]{l}%
j,~\ \ \ \ \ \ \ \mathrm{if\ }\zeta=-1\\
j+1,~\ \mathrm{if\ }\zeta=1
\end{array}
\right.  . \label{a.1}%
\end{align}
The equation is complemented by the condition $E\in(-m,m)$ and the boundary
condition $f_{in}(0)=0$. We note that $\nu$ takes positive half-integer values
as well as $j$ does. The substitution
\begin{equation}
~f_{in}(r)=\sqrt{r}w(z),~z=\eta r \label{a.7}%
\end{equation}
reduces eq. (\ref{a.1}) to the well-known Bessel equation, see \cite{BatEr},%
\begin{equation}
\frac{d^{2}w(z)}{d^{2}z}+\frac{1}{z}\frac{dw(z)}{dz}+\left(  1-\frac{\nu^{2}%
}{z^{2}}\right)  w(z)=0, \label{a.8}%
\end{equation}
complemented by the boundary condition $\sqrt{z}w(z)\rightarrow0$ as
$z\rightarrow0$. The general solution of Eq. (\ref{a.8}) with $\nu\geq1/2$
under this boundary condition is $w(z)=cJ_{\nu}(z)$, where $J_{\nu}(z)$ is the
well-known Bessel function, see \cite{BatEr}. Using the representation for the
function $g_{in}(r)$ in the first row in (\ref{s.9}) and the relation $J_{\nu
}^{\prime}(z)+(\zeta\nu)/z)J_{\nu}(z)=\zeta J_{\nu-\zeta}(z)$, see
\cite{BatEr}, we obtain that the general solution of system (\ref{s.9}%
),(\ref{s.2}) in the region $0\leq r\leq r_{0}$ and under the above-mentioned
conditions is given by%
\begin{align}
f_{in}(r)  &  =c\sqrt{r}J_{\nu}(\eta r)=c\sqrt{r}\left\{
\begin{array}
[c]{c}%
J_{j}(\eta r),~\zeta=-1\\
J_{j+1}(\eta r),~\zeta=1
\end{array}
\right.  ,\nonumber\\
g_{in}(r)  &  =c\sqrt{r}\sqrt{\frac{k_{-}}{k_{+}}}\zeta J_{\nu-\zeta
}(z)=c\sqrt{r}\sqrt{\frac{k_{-}}{k_{+}}}\left\{
\begin{array}
[c]{c}%
-J_{j+1}(\eta r),~\zeta=-1\\
J_{j}(\eta r),~\zeta=1
\end{array}
\right.  . \label{a.11}%
\end{align}
The formulas (\ref{a.11}) give two forms of representation for the functions
$f_{\mathrm{in}}(r)$ and $g_{\mathrm{in}}(r)$: the condensed form in terms of
$\nu,\zeta$ and the expanded form in terms of $j,\zeta=-1$ and\ $j,\zeta=1$.

\subsection{Solving radial equations in region $r_{0}\leq r<\infty$}

In the external region $r_{0}\leq r<\infty$ , where we set
$f(r)=f_{\mathrm{out}}(r)$, $g(r)=g_{\mathrm{out}}(r)$, system (\ref{s.9}%
),(\ref{s.2}) with $E\in(-m,m)$ coincides with the system of the point Coulomb
problem. Solutions of such a system are well-known, see, for example
\cite{book}. The general square-integrable at infinity solution of this system
is given by%

\begin{align}
f_{out}(r)  &  =B\frac{m}{\sqrt{m-E}}(2\beta r)^{\mu}e^{-\beta r}[b_{-}%
\Psi(a+1,c;2\beta r)+\Psi(a,c;2\beta r)],\nonumber\\
g_{out}(r)  &  =B\frac{m}{\sqrt{m+E}}(2\beta r)^{\mu}e^{-\beta r}[b_{-}%
\Psi(a+1,c;2\beta r)-\Psi(a,c;2\beta r)], \label{s5}%
\end{align}
where%
\begin{equation}
\beta=\sqrt{m^{2}-E^{2}},~\mu=\sqrt{\varkappa^{2}-q^{2}},~a=\mu-\frac
{qE}{\beta},\ c=1+2\mu,~b_{-}=\varkappa+\frac{qm}{\beta}, \label{s6}%
\end{equation}
and $\Psi$ is a symbol of one of the standard confluent hypergeometric
functions which vanishes at infinity (it is sometimes called the Tricomi function).

We recall, that there are two standard confluent hypergeometric functions
$\Phi(a,c;x)$ and $\Psi(a,c;x),$ the linearly independent solutions of the
confluent hypergeometric equation, see \cite{BatEr1},%

\begin{align}
\Phi(a,c;x)  &  =%
%TCIMACRO{\dsum \limits_{k=0}^{\infty}}%
%BeginExpansion
{\displaystyle\sum\limits_{k=0}^{\infty}}
%EndExpansion
\frac{(a)_{k}}{(c)_{k}}\frac{x^{k}}{k!},~(a)_{k}=\frac{\Gamma(a+k)}{\Gamma
(a)},~c\notin\mathbb{Z}_{-},\nonumber\\
\Psi(a,c;x)  &  =\frac{\Gamma(1-c)}{\Gamma(a-c+1)}\Phi(a,c;x)+\frac
{\Gamma(c-1)}{\Gamma(a)}x^{1-c}\Phi(a-c+1,2-c;x). \label{r2}%
\end{align}
Rewritten in terms of the Whittaker functions $W$,
\begin{equation}
W_{\lambda,\mu}(x)=e^{-x/2}x^{c/2}\Psi(a,c;x),\ \lambda=\frac{c}{2}%
-a,~\mu=\frac{c}{2}-\frac{1}{2},\nonumber
\end{equation}
see \cite{BatEr1}, solution (\ref{s5}),(\ref{s6}) takes the form
\begin{align}
f_{out}(r)  &  =\frac{B~m}{\sqrt{m-E}}(2\beta r)^{-1/2}[b_{-}W_{\lambda
^{\prime},\mu}(2\beta r)+W_{\lambda,\mu}(2\beta r)],\nonumber\\
g_{out}(r)  &  =\frac{B~m}{\sqrt{m+E}}(2\beta r)^{-1/2}[b_{-}W_{\lambda
^{\prime},\mu}(2\beta r)-W_{\lambda,\mu}(2\beta r)],\ \lambda^{^{\prime}%
}=\frac{qE}{\beta}-\frac{1}{2},\ \lambda=\lambda^{^{\prime}}+1. \label{s7}%
\end{align}

Introducing a new energy variable $\varepsilon$ by $E=m\cos\varepsilon
,~\varepsilon=\arccos\frac{E}{m}\in(0,\pi),$ we rewrite Eqs. (\ref{s7}) as
\begin{align}
f_{out}(r)  &  =B\csc\left(  \frac{\varepsilon}{2}\right)  (2\beta
r)^{-1/2}[(q\csc\varepsilon+\varkappa)W_{\lambda^{\prime},\mu}(2\beta
r)+W_{\lambda,\mu}(2\beta r)],\nonumber\\
g_{out}(r)  &  =B\sec\left(  \frac{\varepsilon}{2}\right)  (2\beta
r)^{-1/2}[(q\csc\varepsilon+\varkappa)W_{\lambda^{\prime},\mu}(2\beta
r)-W_{\lambda,\mu}(2\beta r)], \label{s9a}%
\end{align}
with%
\begin{equation}
\beta=m\sin\varepsilon,\ \lambda=q\cot\varepsilon+\frac{1}{2},~\lambda
^{\prime}=q\cot\varepsilon-\frac{1}{2},~ \label{s(aa}%
\end{equation}
which we take as the final form for the solution of system (\ref{s.9}),
(\ref{s.2}) in the region $r\geq r_{0}$.

\subsection{Continuity conditions and spectrum equation}

After the general solution of system (\ref{s.9}), (\ref{s.2}) is found
independently in the respective regions $0\leq r\leq r_{0}$ and $r_{0}\leq
r<\infty$, it remains to satisfy the basic continuity condition for the
solution as a whole (to sew the partial solutions together smoothly), which
reduces to the requirement of continuity of the solution at the point
$r=r_{0}$:%
\begin{equation}
f_{\mathrm{in}}(r_{0})=f_{\mathrm{out}}(r_{0}),\ \ g_{\mathrm{in}}%
(r_{0})=g_{\mathrm{out}}(r_{0}). \label{s.13}%
\end{equation}

The compatibility of these conditions with $c\neq0$ and $B\neq0$ yields the
transcendental equation, which determines the discrete energy spectrum in the
interval $-m<E<m$ in terms of the variable $\varepsilon$,~$E=m\cos\varepsilon
$,$\ k_{\pm}=m(\cos\varepsilon\pm1)+\frac{q}{r_{0}},$%
\begin{align}
&  \ J_{\nu}(\eta r_{0})\sec\left(  \frac{\varepsilon}{2}\right)  \left[
(\varkappa+q\csc\varepsilon)W_{\lambda^{\prime},\mu}(2\beta r_{0}%
)-W_{\lambda,\mu}(2\beta r_{0})\right] \nonumber\\
&  -\sqrt{\frac{k_{-}}{k_{+}}}\zeta J_{\nu-\zeta}(\eta r_{0})\csc\left(
\frac{\varepsilon}{2}\right)  \left[  (q\csc\varepsilon+\varkappa
)W_{\lambda^{\prime},\mu}(2\beta r_{0})+W_{\lambda,\mu}(2\beta r_{0})\right]
=0. \label{s.14}%
\end{align}

We call this basic equation the spectrum equation for the interval $(-m,m)$.
Strictly speaking, we deal with \ a series of exact spectrum equations for
given $Z,\nu$ and $\zeta$.

It is evident from (\ref{s.14}) that a cutoff removes the degeneracy of the
discrete spectrum in $\zeta$, which is characteristic for a point charge.
After the spectrum equation is solved, the corresponding bound states are
obtained by substituting the evaluated values of bound state energies
$E_{n}(Z,j,\zeta)$ for $E$ in the respective (\ref{a.11}) and (\ref{s5}%
),(\ref{s6}) with due regard to continuity condition (\ref{s.13}), according
to which only normalization factors of the wave eigenfunctions (dublets )
remain undetermined. An analytical solution of the spectrum equation
(\ref{s.14}) with any $Z,\nu,\zeta$ is beyond the scope of our possibilities.

It seems that only numerical solution of these equations is realizable at present.

An equivalent expanded form of the spectrum equation (\ref{s.14}), maybe more
suitable for numerical calculations, is
\begin{align}
\sqrt{\frac{k_{-}}{k_{+}}}\frac{J_{j+1}(\eta r_{0})}{J_{j}(\eta r_{0})}%
+\tan\left(  \frac{\varepsilon}{2}\right)  \frac{\left(  q\csc\varepsilon
-j-\frac{1}{2}\right)  W_{\lambda^{\prime},\mu}(2\beta r_{0})-W_{\lambda,\mu
}(2\beta r_{0})}{\left(  q\csc\varepsilon-j-\frac{1}{2}\right)  W_{\lambda
^{\prime},\mu}(2\beta r_{0})+W_{\lambda,\mu}(2\beta r_{0})}  &  =0,~\zeta
=-1,\nonumber\\
\sqrt{\frac{k_{-}}{k_{+}}}\frac{J_{j}(\eta r_{0})}{J_{j+1}(\eta r_{0})}%
-\tan\left(  \frac{\varepsilon}{2}\right)  \frac{\left(  q\csc\varepsilon
+j+\frac{1}{2}\right)  W_{\lambda^{\prime},\mu}(2\beta r_{0})-W_{\lambda,\mu
}(2\beta r_{0})}{\left(  q\csc\varepsilon+j+\frac{1}{2}\right)  W_{\lambda
^{\prime},\mu}(2\beta r_{0})+W_{\lambda,\mu}(2\beta r_{0})}  &  =0,~\zeta=1.
\label{s14b}%
\end{align}

What concerns a qualitative analysis of the spectrum equation, we can say no
more than the following. It can be shown that the l.h.s. of the spectrum
equation (\ref{s.14}) infinitely oscillates around zero as $E\rightarrow m$,
i.e. $\beta\rightarrow0$ (a proof of the statement appears to be rather
nontrivial). This implies that there exists an infinite set $\{E_{n}%
(Z,j,\zeta)\}$ of roots of the spectrum equation, bound-state energies, with
any fixed $Z,j,\zeta$, which are accumulated at the point $E=m$,
$E_{n}(Z,j,\zeta)\rightarrow m$ as $n\rightarrow\infty$. If $qmr_{0}\ll1$ and
$\mu>0$, the asymptotic behavior of the binding energy $\varepsilon
_{n}=m-E_{n}$ as $n\rightarrow\infty$ is given by%
\begin{equation}
\varepsilon_{n}(Z,j,\zeta)=\frac{q^{2}}{2\left(  n+\mu+\Delta(Z,j,\zeta
)\right)  ^{2}},\text{ \ }\Delta(Z,j,\zeta;r_{0})=(qmr_{0})^{2\mu}%
c(Z,j,\zeta),\text{ }n\rightarrow\infty, \label{s15a}%
\end{equation}
or roughly speaking, $\varepsilon_{n}=\left(  q^{2}n^{-2}/2\right)  \left[
1+O\left(  n^{-1}\right)  \right]  $, $n\rightarrow\infty,$ which reproduces
the well-known result for a nonrelativistic electron in the Coulomb field of a
point charge. In particular, the Zommerfeld spectrum is restored in the limit
$r_{0}\rightarrow0$.

In this paper, we restrict ourselves to the special case of $j=1/2$ and
$\zeta=-1$, which produces the lowest energy levels. In this case, we have%

\begin{equation}
J_{1/2}(z)=\sqrt{\frac{2}{\pi z}}\sin z,\ \mu=\sqrt{1-q^{2}},\ J_{3/2}%
(z)=\sqrt{\frac{2}{\pi z}}\left(  -\cos z+\frac{\sin z}{z}\right)  ,
\label{s16}%
\end{equation}
and the spectrum equation in form (\ref{s14b}) becomes%
\begin{equation}
\sqrt{\frac{k_{-}}{k_{+}}}\left(  \cot\eta r_{0}-\frac{1}{\eta r_{0}}\right)
-\tan\left(  \frac{\varepsilon}{2}\right)  \frac{(q\csc\varepsilon
-1)W_{\lambda^{\prime},\mu}(2\beta r_{0})-W_{\lambda,\mu}(2\beta r_{0}%
)}{(q\csc\varepsilon-1)W_{\lambda^{\prime},\mu}(2\beta r_{0})+W_{\lambda,\mu
}(2\beta r_{0})}=0. \label{s17}%
\end{equation}

We solve this spectrum equation numerically for a series of $Z$ assuming that
the cutoff radius $r_{0}$ is $Z$ dependent according to Eq. (\ref{s.18}).
Results of the numerical calculations are presented in FIG. 2.%

%TCIMACRO{\FRAME{ftbpFU}{2.8046in}{1.9726in}{0pt}{\Qcb{$Z$ dependence of the
%lowest energy levels. For comparison, Zommerfeld energies are indicated by
%circles.}}{}{fig2.eps}{\special{ language "Scientific Word";  type "GRAPHIC";
%width 2.8046in;  height 1.9726in;  depth 0pt;  original-width 11.3057in;
%original-height 7.9321in;  cropleft "0";  croptop "1";  cropright "1";
%cropbottom "0";  filename 'fig2.eps';file-properties "NPEU";}} }%
%BeginExpansion
\begin{figure}[ptb]%
\centering
\includegraphics[
height=1.9726in,
width=2.8046in
]%
{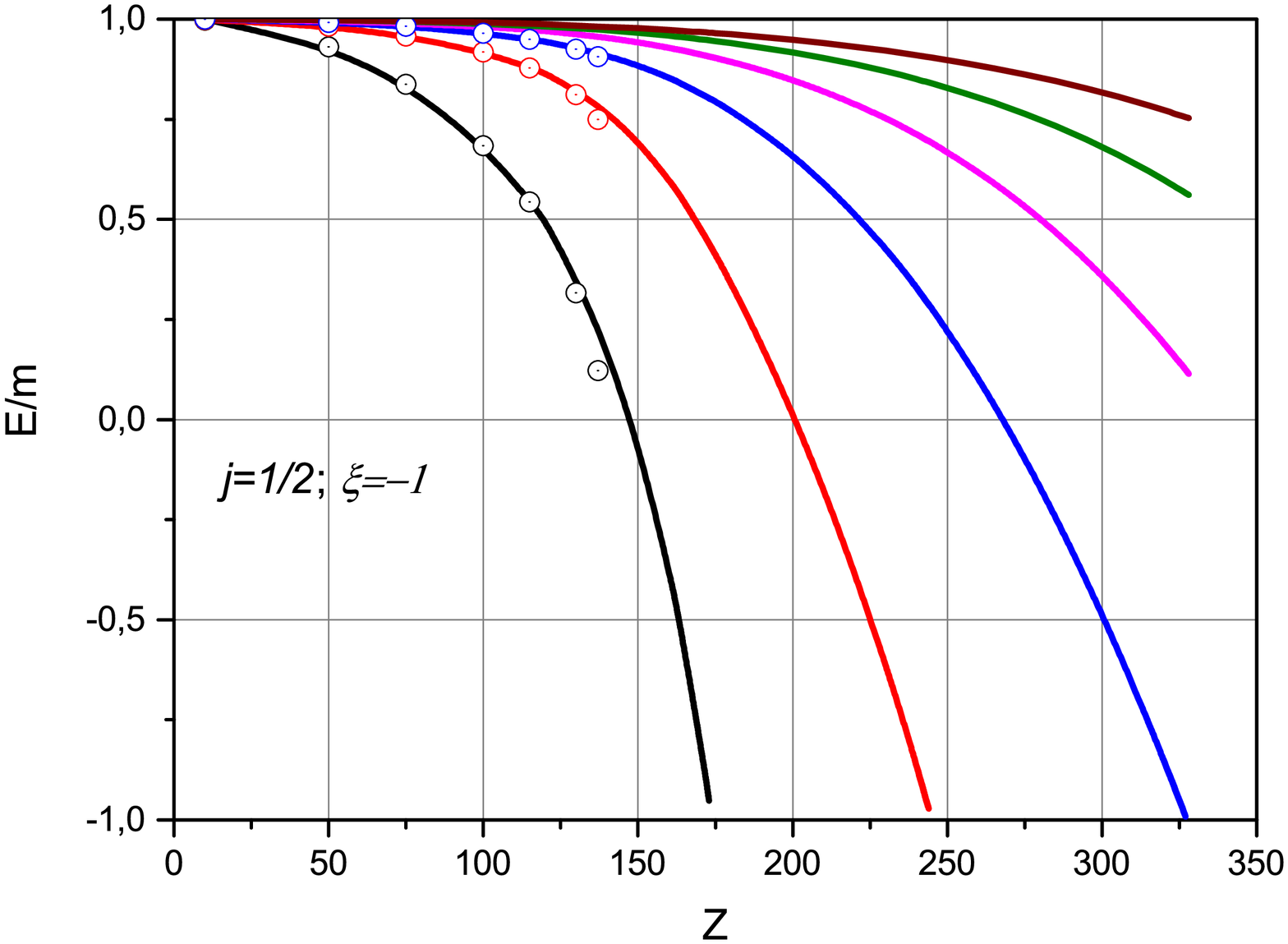}%
\caption{$Z$ dependence of the lowest energy levels. For comparison,
Zommerfeld energies are indicated by circles.}%
\end{figure}
%EndExpansion

\begin{center}%
\begin{tabular}
[c]{|c|c|c|c|}\hline
\multicolumn{4}{|c|}{Table 1. Some numerical data for FIG. 2}\\\hline
$Z$ & $E_{0}$ & $E_{1}$ & $E_{2}$\\\hline
$170$ & $-0,80243$ & $0,47917$ & $0,81528$\\\hline
$172$ & $-0,90118$ & $0,45281$ & $0,80684$\\\hline
$173$ & $-0,95226$ & $0,43936$ & $0,80252$\\\hline
$174$ &  & $0,42573$ & $0,79813$\\\hline
$175$ &  & $0,41192$ & $0,79367$\\\hline
$\vdots$ &  & $\vdots$ & $\vdots$\\\hline
$243$ &  & $-0,94493$ & $0,29796$\\\hline
$244$ &  & $-0,97215$ & $0,28732$\\\hline
$245$ &  &  & $0,27657$\\\hline
$255$ &  &  & $0,16315$\\\hline
$\vdots$ &  &  & $\vdots$\\\hline
\end{tabular}

\end{center}

\section{On bound states with $E=\pm m$,\ supercritical charge\label{S5}}

The preceding consideration is not directly applicable to the points $E=m$ and
$E=-m$. The reason is that formulas (\ref{s5}),(\ref{s6}) break down at these
points because of vanishing the variable $\beta=\sqrt{m^{2}-E^{2}}$ and the
respective blowing up of the factors\ $\frac{1}{\sqrt{m-E}}$ and $\frac
{1}{\sqrt{m+E}}$,$\ $ $b_{-}=\varkappa+\frac{qm}{\beta}$ and the parameter
$a=\mu-\frac{qE}{\beta}$. Each of these points requires a separate consideration.

\subsection{Point $E=m$}

Although it seems evident that there is no bound state with energy $E=m$, for
completeness, we consider this point and show that what seems evident really
holds (actually, an absence of bound states with zero binding energy for an
electron in the attractive Coulomb field is by no means a trivial fact, see a
discussion in the end of the subsection). For this purpose, it is sufficient
to consider system of radial equations (\ref{s.9}) and (\ref{s.2}) for bound
states with $E=m$ in the external region $r\geq r_{0}$ where the system
becomes%
\begin{align}
&  \frac{df(r)}{dr}+\frac{\varkappa}{r}f(r)-\left[  2m+\frac{q}{r}\right]
g(r)=0,\nonumber\\
&  \frac{dg(r)}{dr}-\frac{\varkappa}{r}g(r)+\frac{q}{r}f(r)=0\Longrightarrow
f(r)=\frac{1}{q}\left[  -r\frac{dg(r)}{dr}+\varkappa g(r)\right]
.\label{s1.1}%
\end{align}
It is complemented by the conditions that the both functions $f(r)$ and $g(r)$
are absolutely continuous and square integrable together with their
derivatives on $(r_{0},\infty)$.

Substituting the representation for the function $f(r)$ from the second
row\ in (\ref{s1.1}) into the first equation in (\ref{s1.1}), we obtain that
the function $~g(r)$ satisfies the second-order differential equation
\begin{equation}
r\frac{d^{2}g(r)}{d^{2}r}+\frac{dg(r)}{dr}+2qmg(r)-\frac{\varkappa^{2}-q^{2}%
}{r}g(r)=0.\label{s1.4}%
\end{equation}
The substitution $g(r)=w(z)$,$\ \ z=2\sqrt{2qmr}$, reduces Eq. (\ref{s1.4})
to\ Bessel equation (\ref{a.8}) with $\ \nu=\tilde{\nu}=2\mu$,$\ 2\sqrt
{2qmr_{0}}\leq z<\infty.$ The general solution of this equation is given by
\begin{equation}
w(z)=c_{1}H_{\tilde{\nu}}^{(1)}(z)+c_{2}H_{\tilde{\nu}}^{(2)}(z),\label{s1.7}%
\end{equation}
where $H_{\nu}^{(1)}(z)$ and $H_{\nu}^{(2)}(z)$ are the respective first and
second Hankel functions, see \cite{BatEr1}. Its asymptotic behavior at
infinity is given by%
\begin{align}
&  w(z)=c_{1}\sqrt{\frac{2}{\pi z}}\exp\left[  \frac{i}{4}\left(
4z-2\pi\tilde{\nu}-\pi\right)  \right]  \left[  1+O\left(  \frac{1}{z}\right)
\right]  \nonumber\\
&  \ +c_{2}\sqrt{\frac{2}{\pi z}}\exp\left[  -\frac{i}{4}\left(  4z-2\pi
\tilde{\nu}-\pi\right)  \right]  \left[  1+O\left(  \frac{1}{z}\right)
\right]  ,~z\rightarrow\infty,\label{s1.8a}%
\end{align}
see \cite{BatEr1}. It follows that the asymptotic behavior of the both
functions $f(r)$ and $g(r)$ at infinity is estimated as $f(r),g(r)=O\left(
r^{-1/4}\right)  $,$~r\rightarrow\infty,$ so that the both functions are not
square integrable at infinity. This means that system (\ref{s1.1}) has no
square-integrable solutions, and therefore, there are no bound states with
energy $E=m$, i.e., with zero binding energy, for an electron in the Coulomb
field of any charge $Z$ with cutoff (\ref{s.2}), as well as in the Coulomb
field of a point charge.

The nature of this phenomenon is a long-range character of the Coulomb
potential, which generates an infinite set of bound states with energy levels
accumulated at the point $E=m$, but not reaching this point. This picture is
stable under changing the charge: all these levels go down with increasing
$Z$, but no bound state with zero binding energy appears. A completely
different type of situation occurs in the case of short-range attractive
potentials which can generate bound states with zero binding energy. For
example, an electron in an attractive electric square-well potential can have
such states under certain relations between the radius $r_{0}$ of the well and
its depth $V_{0}$. For $\zeta=1$ and for any $j,$ these relations look rather
simple being given by $\sqrt{V_{0}(2m+V_{0})}r_{0}=z_{n}(j)$,$~n\in\mathbb{N}%
$, where $z_{n}(j)$ are zeroes of the Bessel functions, $J_{j}(z_{n}(j))=0$.
As is well known, these $z_{n}(j)$ form an infinite sequence going to infinity
almost periodically with increasing $n$, $z_{n}(j)\rightarrow\infty$ as
$n\rightarrow\infty$, $z_{n+1}(j)-z_{n}(j)\rightarrow\pi$; in particular,
$z_{n}(1/2)=n\pi$. Accordingly, at fixed radius $r_{0}$, the \ bound states
with given angular momentum $j$ and zero binding energy appears sequentially
and almost periodically with incrasing depth $V_{0}$.

\subsection{Point $E=-m$}

The system of radial equations (\ref{s.9}) and (\ref{s.2}) for bound states
with energy $E=-m$, i.e., with binding energy $2m$, of a relativistic electron
in the Coulomb field with cutoff radius (\ref{s.18}) becomes%
\begin{equation}
\frac{df(r)}{dr}+\frac{\varkappa}{r}f(r)+V(r)g(r)=0,\ \ \frac{dg(r)}{dr}%
-\frac{\varkappa}{r}g(r)-[2m+V(r)]f(r)=0,\label{s2.1}%
\end{equation}
it is complemented by the conditions that the both functions $f(r)$ and $g(r)$
are absolutely continuous together with their first derivatives and
square-integrable on $(0,\infty)$ and are vanishing at zero, $f(0)=$ $g(0)=0$.

\subsubsection{Solving radial equations in region $0\leq r\leq r_{0}$}

The general solution of eqs. (\ref{s2.1}) with (\ref{s.18}) in the internal
region $0\leq r\leq r_{0}$, where we set $f(r)=f_{\mathrm{in}}%
(r),~g(r)=g_{\mathrm{in}}(r)$, under the above-mentioned conditions is given by%

\begin{align}
f_{in}(r)  &  =c\sqrt{r}J_{\nu}(\eta_{0}r)=c\sqrt{r}\left\{
\begin{array}
[c]{c}%
J_{j}(\eta_{0}r),~\zeta=-1\\
J_{j+1}(\eta_{0}r),~\zeta=1
\end{array}
\right.  ,\ \ \eta_{0}=\frac{q}{r_{0}}\sqrt{1-\frac{2mr_{0}}{q}},\nonumber\\
g_{in}(r)  &  =c\sqrt{r}\sqrt{1-\frac{2mr_{0}}{q}}\zeta J_{\nu-\zeta}(\eta
_{0}r)=c\sqrt{r}\sqrt{1-\frac{2mr_{0}}{q}}\left\{
\begin{array}
[c]{l}%
-J_{j+1}(\eta_{0}r),~\zeta=-1\\
J_{j}(\eta_{0}r),~\zeta=1
\end{array}
\right.  , \label{s2.3}%
\end{align}

it is sufficient to put $E=-m$ and $r_{0}=r_{0}(Z)$ in (\ref{a.11}).

\subsubsection{Solving radial equations in region $r_{0}\leq r<\infty$}

In the external region $r_{0}\leq r<\infty$, where we set $f(r)=f_{out}(r)$,
$g(r)=g_{out}(r)$, system (\ref{s2.1}) becomes%
\begin{align}
&  \frac{df_{out}(r)}{dr}+\frac{\varkappa}{r}f_{out}(r)-\frac{q}{r}%
g_{out}(r)=0\Rightarrow g_{out}(r)=\frac{1}{q}\left[  r\frac{df_{out}(r)}%
{dr}+\varkappa f_{out}(r)\right]  ,\nonumber\\
&  \frac{dg_{out}(r)}{dr}-\frac{\varkappa}{r}g_{out}(r)+\frac{q}{r}%
f_{out}(r)-2mf_{out}(r)=0,\label{s2.5}%
\end{align}
it is complemented by the conditions that the both functions $f_{out}(r)$ and
$g_{out}(r)$ are absolutely continuous and square integrable together with
their derivatives on $(r_{0},\infty)$.

Substituting the representation for the function $g_{out}(r)$ from the first
row in (\ref{s2.5}) into the second equation in (\ref{s2.5}), we obtain that
the function $~f_{out}(r)$ satisfies the second-order differential equation
\begin{equation}
r\frac{d^{2}~f_{out}(r)}{d^{2}r}+\frac{d~f_{out}(r)}{dr}-2qm~f_{out}%
(r)-\frac{\varkappa^{2}-q^{2}}{r}~f_{out}(r)=0.\label{s2.7}%
\end{equation}
The substitution $~f_{out}(r)=w(z),~z=2\sqrt{2qmr}$, reduces eq. (\ref{s2.7})
to the equation for the modified Bessel functions (Bessel functions of pure
imaginary argument), see \cite{BatEr}:%
\begin{equation}
\frac{d^{2}w(z)}{d^{2}z}+\frac{1}{z}\frac{dw(z)}{dz}-\left(  1+\frac
{\tilde{\nu}^{2}}{z^{2}}\right)  w(z)=0,\ \tilde{\nu}=2\mu,~2\sqrt{2qmr_{0}%
}\leq z<\infty.\label{s2.9}%
\end{equation}
\ 

The requirement for $~f_{out}(r)$ to be square-integrable at infinity then
yields $~f_{out}(r)=AK_{\tilde{\nu}}(z),$ where $K_{\tilde{\nu}}(z)~$is the
MacDonald function,%
\begin{align}
K_{\tilde{\nu}}(z)  &  =\frac{\pi}{2\sin\pi\tilde{\nu}}\left[  I_{-\tilde{\nu
}}(z)-I_{\tilde{\nu}}(z)\right]  ,\ \ \tilde{\nu}\neq n\in\mathbb{Z}%
_{+},\nonumber\\
I_{\tilde{\nu}}(z)  &  =%
%TCIMACRO{\dsum \limits_{m=0}^{\infty}}%
%BeginExpansion
{\displaystyle\sum\limits_{m=0}^{\infty}}
%EndExpansion
\frac{(z/2)^{2m+\tilde{\nu}}}{m!\Gamma(m+\tilde{\nu}+1)},\ \ K_{\tilde{\nu}%
}(z)=K_{-\tilde{\nu}}(z). \label{s2.11}%
\end{align}

For $\tilde{\nu}=n\in\mathbb{Z}_{+}$, the functions $K_{n}(z)$ contain terms
with a logarithmic factor, see \cite{BatEr}.

Using $f_{out}(r)=AK_{\tilde{\nu}}(z)$ and the representation for the function
$g_{out}(r)$ in the first row in (\ref{s2.5}), we finally obtain that the
general solution of system (\ref{s2.5}) under the above-mentioned conditions
is given by%
\begin{align}
&  \ f_{out}(r)=AK_{\tilde{\nu}}(z),\ \ g_{out}(r)=\frac{A}{q}\left[  \frac
{z}{2}\frac{d}{dz}K_{\tilde{\nu}}(z)+\varkappa K_{\tilde{\nu}}(z)\right]
\nonumber\\
&  =\frac{A}{q}\left\{  -\frac{z}{4}\left[  K_{\tilde{\nu}-1}(z)+K_{\tilde
{\nu}+1}(z)\right]  +\varkappa K_{\tilde{\nu}}(z)\right\}  ,\ \text{ }%
z=2\sqrt{2qmr}, \label{s2.12}%
\end{align}
where we use the known formula $K_{\tilde{\nu}-1}(z)+K_{\tilde{\nu}%
+1}(z)=-2K_{\tilde{\nu}}^{\prime}(z)$ (see \cite{BatEr}).

\subsubsection{Charges providing bound states with energy $E=-m$,
supercritical charge}

After the general solution of system (\ref{s2.1}) is found independently in
the respective regions $0\leq r\leq r_{0}$ and $r_{0}\leq r<\infty$, it
remains to satisfy the basic continuity condition for the solution as a whole
(to sew the partial solutions together smoothly), which reduces to the
requirement of continuity of the solution at the point $r=r_{0}\left(
Z\right)  $,%
\begin{equation}
f_{in}(r_{0})=f_{out}(r_{0}),\ \ g_{in}(r_{0})=g_{out}(r_{0}). \label{s2.13}%
\end{equation}

The compatibility of equalities (\ref{s2.13}) with $c\neq0,~A\neq0$ yields the
relation%
\begin{align}
&  J_{\nu}(\eta_{0}r_{0})\left\{  -\frac{z_{0}}{4}\left[  K_{\tilde{\nu}%
-1}(z_{0})+K_{\tilde{\nu}+1}(z_{0})\right]  +\varkappa K_{\tilde{\nu}}%
(z_{0})\right\}  -(\eta_{0}r_{0})\zeta J_{\nu-\zeta}(\eta_{0}r_{0}%
)K_{\tilde{\nu}}(z_{0})=0,\ \nonumber\\
\  &  \eta_{0}r_{0}=q\sqrt{1-\frac{2mr_{0}}{q}},\ \ z_{0}=2\sqrt{2qmr_{0}%
},~\nu=j+\frac{\zeta+1}{2},~\tilde{\nu}=2\sqrt{\left(  j+\frac{1}{2}\right)
^{2}-q^{2}},\ \varkappa=\zeta\left(  j+\frac{1}{2}\right)  , \label{s2.14}%
\end{align}
that can be considered as the (transcendental) equation for charges $Z$
providing bound states with energy $E=-m$ for an electron with given total
angular momentum $j$ and spin number $\zeta$. We let $Z^{(-m)}(j,\zeta)$
denote such charges.

An analytical solution of equation (\ref{s2.14}) for $Z^{(-m)}(j,\zeta)$ with
arbitrary $j$ and $\zeta$ is unlikely to be possible at present. We only can
try to analize it qualitatively and solve it numerically.

An equivalent form of equation (\ref{s2.14}) that seems more suitable for its
qualitative analysis and its numerical solution is the equation
\begin{equation}
(\eta_{0}r_{0})\zeta\frac{J_{\nu-\zeta}(\eta_{0}r_{0})}{J_{\nu}(\eta_{0}%
r_{0})}+\left[  \frac{z_{0}}{4}\frac{K_{\tilde{\nu}-1}(z_{0})+K_{\tilde{\nu
}+1}(z_{0})}{K_{\tilde{\nu}}(z_{0})}-\varkappa\right]  =0. \label{s2.14a}%
\end{equation}

What concerns a qualitative analysis of Eq. (\ref{s2.14a}), we can say the
following.\ As follows from the well-known asymptotic behavior of the Bessel
function $J_{\nu}(z)$ as $z\rightarrow\infty$, the ratio of the Bessel
functions $J$ multiplied by $\zeta$ in the l.h.s. of (\ref{s2.14a}), $\zeta
J_{\nu-\zeta}(\eta_{0}r_{0})/J_{\nu}(\eta_{0}r_{0})$, oscillates with
increasing $Z$ around zero almost periodically, the period is $\pi\alpha^{-1}%
$, and ranges from $\infty$ to $-\infty$ (more specifically, as $\cot
[q+\pi/4(1-2j)]$ for $\zeta=-1$ and as $-\tan[q+\pi/4(1-2j)]$ for $\zeta=1$).\ 

A\ plausible estimate of the asymptotic behavior of the MacDonald function
$K_{i\sigma}(a\sqrt{\sigma}),\ a,\sigma\in\mathbb{R}$, as $\sigma
\rightarrow\infty$ allows a conclusion that the behavior of the ratio of the
MacDonald functions $K$ in the l.h.s. of (\ref{s2.14a}) with increasing $Z$ is
similar, a difference is that the oscillation frequency grows logarithmically
with $Z$. It follows that each Eq. (\ref{s2.14a}) with any fixed $j,\zeta$ has
an infinite sequence $\{Z_{n}^{(-m)}(j,\zeta),n\in\mathbb{N}\}$ of solutions,
$Z_{n}^{(-m)}(j,\zeta)\rightarrow\infty$ as $n\rightarrow\infty$, and the
difference between the subsequent terms in this sequence decreases with $n$.

We are now going to discuss the notion of the so-called supercritical
charge.It seems that at present, there is no generally excepted understanding
of this notion among physicists. In particular, each $Z_{n}^{(-m)}(j,\zeta
)$\ of the whole set $\cup_{j,\zeta}\{Z_{n}^{(-m)}(j,\zeta),n\in\mathbb{N}\}$
is sometimes called the supercritical charge (or sometimes critical charge, as
in Refs. \cite{ZelPo72,UPN}), so that there is an infinite set of
supercritical charges of nonclear physical meaning. We cannot agree with such
a viewpoint for at least two reasons. First, almost all $Z_{n}^{(-m)}%
(j,\zeta)$ are nonintegral and therefore have no direct physical meaning.
Second, our standpoint is that the supercritical charge must be unique and
integer valued. It remains to specify the value of supercritical charge. We
believe that the supercritical charge is defined by the minimum among all the
charges $Z_{n}^{(-m)}(j,\zeta)$, which is achieved in the sector
$j=1/2,\ \zeta=-1$ and equals to $Z_{1}(1/2,-1)$ (to our knowledge, many
physicists hold this viewpoint). Namely, we define the supercritical charge
$Z_{\mathrm{scr}}$ as an integer nearest from above to $Z_{1}(1/2,-1)$:%
\begin{equation}
Z_{\mathrm{scr}}=\left\{
\begin{array}
[c]{l}%
\lbrack Z_{1}(1/2,-1)]+1\text{ \textrm{if }}Z_{1}(1/2,-1)\text{ \textrm{is
noninteger}}\\
Z_{1}(1/2,-1)\text{ \textrm{if }}Z_{1}(1/2,-1)\text{ \textrm{is integer}}%
\end{array}
\right.  , \label{s2.14d}%
\end{equation}
the symbol $[...]$ denotes the integral part of a real number. It should be
emphasized that as well as the whole set $\cup_{j,\zeta}\{Z_{n}^{(-m)}%
(j,\zeta),n\in\mathbb{N}\}$, the supercritical charge $Z_{\mathrm{scr}}$
depends on a type of regularization of the Coulomb field, i.e., on the charge
distribution in a nucleus of finite radius. It is different for a uniformly
charged sphere and for a uniformly charged ball. Shortly speaking, the
supercritical charge is model dependent.

Returning to equation (\ref{s2.14a}) for $Z_{n}^{(-m)}(j,\zeta)$ , we restrict
ourselves to the case $j=1/2,\ \zeta=-1$ and, in particular, find the
supercritical charge.

In this case, equation. (\ref{s2.14a}) for $Z^{(-m)}(1/2,-1)$ becomes, see
(\ref{s16}),
\begin{equation}
\varphi\left(  Z\right)  =0,\ \ \varphi\left(  Z\right)  =(\eta_{0}r_{0}%
)\cot(\eta_{0}r_{0})+\frac{z_{0}}{4}\frac{K_{\tilde{\nu}-1}(z_{0}%
)+K_{\tilde{\nu}+1}(z_{0})}{K_{\tilde{\nu}}(z_{0})} \label{s2.15a}%
\end{equation}
with%
\begin{equation}
\eta_{0}r_{0}=q\sqrt{1-\frac{2mr_{0}}{q}},\ \ z_{0}=2\sqrt{2qmr_{0}}%
,~\tilde{\nu}=2\sqrt{1-q^{2}}. \label{s2.15b}%
\end{equation}

In exemining equation (\ref{s2.15a}), we have to distinguish two regions of
$Z$, namely, the region $0<Z\leq Z_{\mathrm{c}}$ $(0<q\leq1)$ and the region
$Z>Z_{\mathrm{c}}$ $(q>1)$. In the first region, the parameter $\tilde{\nu}$
is real-valued, while in the second region, $\tilde{\nu}$ becomes pure
imaginary, $\tilde{\nu}=$ $i\sigma,\ \sigma>0$. The variable $z_{0}$ is
real-valued and positive in the both regions. As is known, the MacDonald
function of positive argument and real index is strictly positive, see
\cite{88}, while the MacDonald function of positive argument and pure
imaginary index is alternating in sign. It follows that in the first region of
$Z$, the second term in Eq. (\ref{s2.15a}) is strictly positive, in fact, with
any $r_{0}$,
\begin{equation}
\frac{z_{0}}{4}\frac{K_{\tilde{\nu}-1}(z_{0})+K_{\tilde{\nu}+1}(z_{0}%
)}{K_{\tilde{\nu}}(z_{0})}>0,\ 0<Z\leq Z_{\mathrm{c}},\ \forall r_{0},
\label{s2.16}%
\end{equation}
while in the second region of $Z$, this term is alternating in sign. On the
other hand, as is easily verified, in the first region of $Z$, the first term
in Eq. (\ref{s2.15a}) is strictly positive, \ in fact, with any $r_{0}$,
\begin{equation}
(\eta_{0}r_{0})\cot(\eta_{0}r_{0})>0,\ 0<Z\leq Z_{\mathrm{c}},\ \forall r_{0}.
\label{s2.17}%
\end{equation}
Really, if $q<2mr_{0}$, we have $\eta_{0}r_{0}=i\tau,\tau>0$, and $(\eta
_{0}r_{0})\cot(\eta_{0}r_{0})=\tau\coth\tau>0$, while if $q>2mr_{0}$, we have
$0<\eta_{0}r_{0}<1$ and $\cot(\eta_{0}r_{0})>\cot1>0$. In the second region,
this term is evidently oscillating function ranging from $\infty$ to $-\infty$
with incrteazing $Z$,
\begin{equation}
(\eta_{0}r_{0})\cot(\eta_{0}r_{0})\in(-\infty,\infty),\ Z\in(Z_{\mathrm{c}%
},\infty). \label{s2.18}%
\end{equation}

It follows from (\ref{s2.16}) and (\ref{s2.17}) that in the first $Z$ region,
$0<Z\leq Z_{\mathrm{c}}$, equation (\ref{s2.15a}) for $Z^{(-m)}(1/2,-1)$ has
no solution, or equivalently, under any $r_{0}$, there is no charge in the
region $0<Z\leq Z_{\mathrm{c}}$ providing bound state with energy $E=-m$. On
the contrary, in the second $Z$ region, $Z>Z_{\mathrm{c}}$, equation
(\ref{s2.15a}) has \ an infinite growing sequence of solutions, as was already
stated above. All this allows us to make the assertion that there is an
infinitely growing \ sequence $\{Z_{n}^{(-m)}(1/2,-1),\ n\in\mathbb{N}%
,\ \ Z_{1}^{(-m)}(1/2,-1)>Z_{\mathrm{c}}\}$ of charges providing the bound
state with energy $E=-m$. It is interesting to compare this situation with a
similar situation in the case of the pure Coulomb field of a point charge
(without a cutoff). We know from \cite{book} that for a pure Coulomb field
with $Z<Z_{\mathrm{s}}=(\sqrt{3}/2)Z_{\mathrm{c}}$, in which case the Dirac
Hamiltonian is defined uniquely, there is no bound state with $E=-m$, while
for each $Z\geq Z_{\mathrm{s}}$, such a bound state does exist. More exactly,
for each $Z\geq Z_{\mathrm{s}}$, the s.a. radial Hamiltonian $\hat
{h}(Z,1/2,-1)$ is defined nonuniquely; instead, there is a one-parameter
family of s.a. Hamiltonians specified by certain s.a. boundary conditions at
zero, and among these, there is a Hamiltonian, specified by peculiar s.a.
boundary conditions at zero, which has a bound state with energy $E=-m$.

We solve Eq. (\ref{s2.15a}) numerically and find several first terms of the
infinite sequence $\{Z_{n}^{(-m)}(1/2,-1)\}$. Results of the numerical
calculations are presented in FIG. 3 and by Eq. (\ref{s2.19})%

\begin{figure}[ptbh]%
\centering
\includegraphics[
height=2.066in,
width=3.1168in
]%
{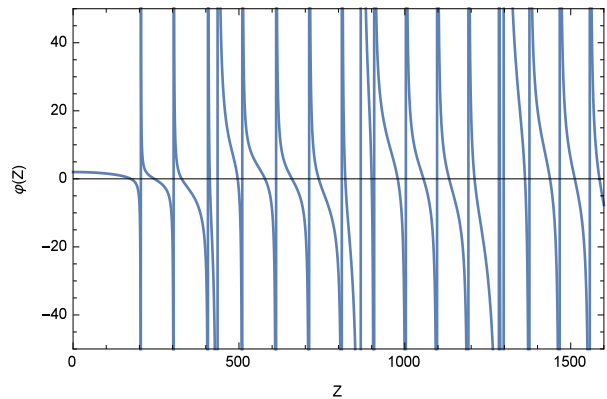}%
\caption{Graphic solution of Eq. (\ref{s2.15a}).}%
\end{figure}
%EndExpansion
%

\begin{align}
&  \{Z_{n}^{(-m)}\left(  1/2,-1\right)
\}=\{173.92;\ 245.01;\ 327.39;\ 412.15;\ 496.11;\ 578.55;\ 659.82;\ 740.44;\ 820.53;\nonumber\\
&
899.93;\ 978.50;\ 1056.28;\ 1133.46;\ 1210.16;\ 1286.43;\ 1362.21;\ 1437.48;\ ...\}.
\label{s2.19}%
\end{align}

We note, that as was already stated above, all the presented terms are
noninteger and have no direct physical meaning. It is also worth noting that
the difference between the subsequent terms of the sequence decreases with
$n$, as was expected. For us, only the first term of the sequence is
important. According to our definition, the supercritical charge in our model
is $Z_{\mathrm{scr}}=174$\textrm{ }with$\ r_{0}=9.47\times10^{-15}%
\mathrm{m}\approx10\mathrm{F}$ which almost coincides with the result by Popov
in \cite{Popov70a}.

\section{Conclusion}

We share an opinion widespread among physicists\ that the supercritical charge
marks a boundary, after which, i.e., for $Z\geq Z_{\mathrm{scr}}$, the
description of a behavior of an electron in strong Coulomb field, even
regularized at the origin, in the framework of one-particle relativistic
quantum mechanics based on the Dirac Hamiltonoian definitely fails. It is
believed that if $Z\geq Z_{\mathrm{scr}}$, processes with a fixed number of
particles do not exist, in particular, do not exist pure one-particle
processes, any process is accompanied by multiple $e^{+}e^{-}$ pair creation.
In such a situation, only a consistent QED with strong Coulomb field may
provide rules for calculating all the quantum processes. The same concerns the
critical charge $Z_{\mathrm{c}}=\alpha^{-1}$, and maybe even the lower
critical charge $Z_{\mathrm{s}}=(\sqrt{3}/2)\alpha^{-1}$, for the Coulomb
field of a point nucleus. It also must be remembered that although
one-particle relativistic quantum mechanics with Dirac Hamiltonian with any
potential, including the Coulomb field of any charge $Z$, point or nonpoint,
is mathematically consistent, in particular, describes a unitary evolution, it
is unsatisfactory from the physical standpoint because of the unboundedness of
the electron energy spectrum from below. As is well known, this drawback is
overecome by secondary Fermi--Dirac quantization and transition to
many-particle QFT. In any case, only the future QED can provide a proper
description of a behavior of an electron in the Coulomb field of any strength.

It should be noted that the existing heavy nuclei can imitate time-dependent
supercritical Coulomb fields at collisions. Then one can try to calculate the
pair creation effect using elements of the well-elaborated QFT with
time-dependent external electric fields that are switched on and off at the
respective initial and final instants of time, see \cite{Grein85,Shaba15} and
citations therein.

\begin{acknowledgement}
Gitman is grateful to the Brazilian foundations FAPESP and CNPq for permanent
support. His work is also partially supported by the Tomsk State University
Competitiveness Improvement Program. The reported study of DMG was partially
supported by RFBR, research project No. 15-02-00293a; Voronov thanks RFBR,
grant 14-02-01171.
\end{acknowledgement}

\end{document}